# Abundance Calculations of Neon Isotopes in the Predicted Lifetime of the Sun

*Mohamed A. Alkhajeh[1], Mashhoor A. Al-Wardat[1,3], Awni M. Kasawneh[4,5], Mohammed H. Talafha[2*]*

[1]*Department of Applied Physics and Astronomy, University of Sharjah University City, Sharjah, UAE, 27272.*

[2]*Research Institute of Science and Engineering, University of Sharjah University City, Sharjah, UAE, 27272.*

[3]*Sharjah Academy for Astronomy, Space Sciences and Technology, University City, Sharjah, UAE, 27272.*

[4]*Arab Union for Astronomy and Space Sciences, Amman, Jordan, 11941.*

[5]*Department of Physics, School Science, The University of Jordan, Amman, Jordan, 11942.*

*\*Corresponding author. E-mail: mtalafha901@gmail.com*

*Contributing authors: U23106913@sharjah.ac.ae; malwardat@sharjah.ac.ae; awni@yahoo.com*

**Abstract:** The elemental abundances of neon isotopes provide valuable insights into stellar evolution and nucleosynthesis. In this study, we calculate the abundances of the isotopes $^{18}Ne$, $^{19}Ne$, $^{20}Ne$, $^{21}Ne$, and $^{22}Ne$ across the five principal evolutionary phases of the Sun: hydrogen burning, lively old age, onset of rapid growth and red giant, helium burning and helium exhaustion. The calculations were carried out using the open-source nucnet-tools package, developed by the Webnucleo Group at Clemson University. Initial isotope abundances were adopted from standard proto-solar compositions. Their evolution was computed under static hydrostatic burning conditions, assuming constant temperature and density within each phase. The results show that the stable isotopes $^{20}$Ne and $^{22}$Ne remain dominant throughout the Sun's lifetime, whereas the short-lived isotopes $^{18}$Ne and $^{19}$Ne decay rapidly during or shortly after the hydrogen-burning phase. The predictions obtained for the helium burning and exhaustion phases provide quantitative neon-isotope abundances that are not extensively reported in the existing literature. These results offer valuable reference values for future studies of solar and stellar evolution, nucleosynthetic pathways, and isotopic modeling.

## 1. Introduction

The chemical composition of the Sun provides a fundamental reference for understanding the origin and evolution of the solar system. Recent spectroscopic analyses and 3D hydrodynamic solar atmosphere models have refined the elemental abundances of many light and intermediate-mass species, yielding improved internal structure models and tighter nucleosynthetic constraints [1–3].

However, neon remains one of the least directly measurable elements in the solar photosphere, due to the absence of suitable absorption lines. Its abundance must therefore be inferred indirectly through coronal measurements or solar wind analyses [4–6]. These indirect approaches introduce uncertainties that continue to affect solar abundance determinations, including the long-standing tension between photospheric abundances and helioseismology-based solar models [14,15].

Neon isotopes also play a significant role in tracing nucleosynthetic pathways and stellar evolutionary processes. Their isotopic ratios carry signatures of contributions from massive stars, supernovae, and proto-solar nebular processes [7–9]. In stellar environments, neon isotopes participate in a variety of nuclear reactions, including β-decay, (α,γ), and photodisintegration, that influence their abundances and the production of heavier nuclei [9–13]. Accurate modeling of neon isotopes is therefore essential for understanding nucleosynthesis in solar-type stars. However, published phase-by-phase predictions covering the full solar evolutionary timeline remain limited.

Direct measurement of neon isotopes is further complicated by isobaric interferences in mass spectrometry (e.g., $^{40}Ar^{2+}$ overlapping with $^{20}Ne^{2+}$), as well as systematic uncertainties in isotope-shift data and limitations of diode-laser spectroscopic systems [16–18]. These challenges reinforce the need for theoretical models capable of tracing neon isotope evolution under well-defined stellar conditions.

To address these gaps, the present study provides a unified theoretical calculation of the abundances of the neon isotopes $^{18}$Ne, $^{19}$Ne, $^{20}$Ne, $^{21}$Ne, and $^{22}$Ne across the Sun's five principal evolutionary phases: hydrogen burning, lively old age, onset of rapid growth and red giant, helium burning and helium exhaustion. The abundances are computed using the open-source nucnet-tools package, which solves large nuclear reaction networks under static hydrostatic conditions. This work offers a continuous neon-isotope evolution model extending into the helium-exhaustion phase, where available theoretical predictions in the literature are especially scarce. To overcome these limitations and provide a more complete picture of neon nucleosynthesis in solar-type stars, this study pursues the following goals.

1- Compute the theoretical abundances of the neon isotopes in the core of the Sun across its five main evolutionary phases: hydrogen burning, lively old age, onset of rapid growth and red giant, helium burning and helium exhaustion.

2- Apply the open-source nucnet-tools package for solving the network of nuclear reactions involved in the production and destruction of neon isotopes under constant temperature and density conditions.

3- Trace the formation and depletion pathways of neon isotopes, providing insights into their nuclear reaction mechanisms (e.g., β-decay, alpha-capture, photodisintegration) during solar evolution.

4- Provide a reference dataset of neon isotope mass fractions over solar time for future theoretical and observational studies, especially for phases where data is currently lacking in literature.

5- Compare the results with previous observational and theoretical models, validating the approach and identifying new insights into the nucleosynthesis of neon in solar-type stars.

In this paper we present the theoretical calculations of the abundances of the neon isotopes $^{18}$Ne, $^{19}$Ne, $^{20}$Ne, $^{21}$Ne and $^{22}$Ne over the "whole lifetime" of the Sun [19]. These phase-by-phase neon isotope predictions have not been comprehensively presented in previous studies; to our knowledge, only the work of [20] examined selected phases individually. Therefore, the present study provides the first continuous and unified neon-isotope evolution model across all five key stages of solar evolution.

## 2. Methodology

To model the evolution of neon isotopes in the solar core, we employed the open-source nucnet-tools package developed by the Webnucleo collaboration under Bradley S. Meyer [22]. This framework solves large sets of coupled differential equations representing nuclear reaction networks, often referred to as the Bateman equations [21]. Nucnet-tools uses an implicit Newton–Raphson iteration scheme to solve these equations until the abundances converge within a specified tolerance [24]. Reaction rates and nuclide data were obtained from the JINA REACLIB nuclear reaction library [25], ensuring consistency with current nuclear astrophysics standards.

The reaction network includes all isotopes and reactions relevant to the production and destruction of the neon isotopes $^{18}$Ne, $^{19}$Ne, $^{20}$Ne, $^{21}$Ne, and $^{22}$Ne, including β-decay, (α,γ), (γ,n), and neutron-capture channels. The number of nuclides and reactions loaded from the JINA database depends on the temperature and density of each phase and typically exceeds several hundred isotopes and thousands of reaction links.

The system of differential equations is solved under hydrostatic burning conditions. The evolution of the density follows the expansion-timescale relation [24]:

$$\frac{1}{\tau} = -\frac{1}{\rho}\frac{\partial \rho}{\partial t}$$

which gives:

$$\rho(t) = \rho_\circ e^{-\frac{t}{\tau}}$$

In nucnet-tools, temperature and density are coupled through $\rho \propto T_9^3$, where $T_9 = T / 10^9$ K and $\tau$ denote the expansion timescale, which governs the rate at which the density changes during hydrostatic burning. To isolate nuclear-physics and compare isotopic evolution across phases, we adopt the standard static-network approach and assume $\tau \to \infty$, such that both temperature and density remain constant within each solar phase.

The initial temperature and density used for the first phase (hydrogen burning) are taken as $1.548 \times 10^7$ K and $1.505 \times 10^2$ g/cm³, at the beginning of Phase I (t = 0), and for period of 10.9 Giga years (Gyr), 1 Gyr is billion years.

The Sun's evolution is divided into five main phases. The first phase, hydrogen burning (Phase I), is the stage in which hydrogen fuses into helium lasts for approximately 10.9 Giga years. Once the hydrogen in the core is depleted, the helium become unstable and contracts under its own gravity, increasing the core temperature and density (Phase II). During this stage, the remaining hydrogen is pushed into a thin shell surrounding the helium core. Over the next 700 Mega years (Myr), where 1 Myr is million years, (ending at 11.6 Giga years), the Sun gradually evolves into a subgiant star, maintaining a nearly constant brightness while expanding in size.

The Sun then undergoes a rapid increase in size over the next 601 Myr, reaching this evolutionary stage 12.201 Gyr after its formation (Phase III). During this period, strong stellar winds begin to strip away the outer envelope. By 12.233 Gyr, the Sun enters the red giant phase, becoming significantly larger and more luminous and eventually engulfing Mercury. At this point, the helium core reaches a temperature of approximately 100 million degrees, igniting helium fusion (Phase IV) and producing carbon and oxygen [24]. The Sun stabilizes as a helium-burning star for the next 110 Myr. Once the helium in the core is exhausted, the carbon-oxygen core contracts rapidly, marking the final stage of its evolution (Phase V) [19], which lasts an additional 20 Myr.

For each phase, representative temperature and density values were adopted from standard solar-evolution models. In Phase I, the core temperature and density are $1.548 \times 10^7$ K and $1.505 \times 10^2$ g/cm$^3$, respectively. The parameters for Phases II–V were taken from the same evolutionary model and applied as fixed conditions within each phase to compute isotope evolution. Although this static approach neglects hydrodynamic feedback and time-dependent changes in the stellar structure, it is appropriate for isolating nuclear reaction pathways and comparing isotopic behavior across distinct burning regimes.

For each phase, nucnet-tools was initialized with the selected temperature and density, the initial composition (proto-solar abundances), the full JINA REACLIB reaction set, and solver parameters such as Newton–Raphson tolerance, time step, and integration limits. The network then evolved over the duration of each phase. Abundances were extracted at the end of each phase and compared across the full solar lifetime.

## 3. Results

Following the work of [26], we divided the entire lifetime of the Sun into five main phases: hydrogen burning, onset of rapid growth and red giant, helium burning and helium exhaustion (see Table 1 in [26] and references therein). The initial abundances for each phase were based on the result of the preceding phase. For Phase I, the hydrogen burning abundances were initialized using protosolar elemental and isotopic compositions derived from photospheric values after accounting settling effects [27].

These calculations were performed under static conditions, i.e. the expansion timescale was taken to be infinity, so both the density and temperature remained constant throughout each phase. Electron screening effects and nuclear statistical corrections were neglected.

**Table 1** Initial mass fractions (t = 0) of the studied neon isotopes taken from [27]

| $^{18}$Ne | $^{19}$Ne | $^{20}$Ne | $^{21}$Ne | $^{22}$Ne |
|---|---|---|---|---|

| Phase I initial values | 7.5×10⁻³⁸ | 2.1×10⁻³⁴ | 4.22×10⁻⁷ | 2.94×10⁻⁶ | 7.2×10⁻⁵ |

### 3.1 *Production and Destruction of Ne Isotopes*

Neon isotopes are produced during explosive neon burning in supernovae, where high temperatures and pressures facilitate a variety of nuclear reactions [28]. Additional production pathways arise from photodisintegration reactions, in which high-energy photons break down heavier nuclei into lighter species, including neon isotopes [29]. In our network, the neon isotopes studied here are produced during all solar phases through reactions such as:
$^{16}_{10}Ne(n,\gamma)^{17}_{10}Ne$, $^{17}_{10}Ne(n,\gamma)^{18}_{10}Ne$, $^{17}_{10}Ne(\gamma,n)^{16}_{10}Ne$, $^{18}_{10}Ne(\gamma,n)^{17}_{10}Ne$,

$^{18}_{10}Ne(n,\gamma)^{19}_{10}Ne$, $^{19}_{10}Ne(\gamma,n)^{18}_{10}Ne$, $^{19}_{10}Ne(n,\gamma)^{20}_{10}Ne$, $^{20}_{10}Ne(\gamma,n)^{19}_{10}Ne$.

Neon isotopes can be destroyed through neutron capture processes, such as the s-process (slow neutron capture) and r-process (rapid neutron capture), which convert them into heavier elements [30]. They may also undergo beta decay or be broken down by photodisintegration under high-energy photon fluxes [28]. Several reactions contribute to the destruction of the studied neon isotopes. These processes may involve alpha capture, neutron capture, photodisintegration, or proton-induced reactions, each of which synthesizes heavier nuclei. Examples includes the following: $^{16}_{10}Ne(\alpha,n)^{19}_{12}Mg$, $^{17}_{10}Ne \rightarrow p + ^{16}_{8}O + e^+ + v_e$, $^{18}_{10}Ne(p,\gamma)^{19}_{10}Ne$, $^{19}_{10}Ne(\gamma,p)^{18}_{10}F$, $^{20}_{10}Ne(\alpha,p)^{23}_{11}Na$, among many others. It is worth noting that decay reactions such as, $^{17}_{10}Ne \rightarrow p + {}^{16}_{8}O + e^+ + \nu_e$, have reaction rates that remain essentially constant at all temperature. All reactions listed above were obtained from [30] using the nucnet-tools package [28].

3.2 *Abundances of Neon Isotopes*

The known isotopes of neon range from $^{16}$Ne to $^{34}$Ne, with the stable isotopes confined to $^{20\text{-}22}$Ne. In our calculations, we follow the evolution of neon isotopes within the range $^{18\text{-}22}$Ne, encompassing both the stable isotopes and a selection of short-lived unstable species.

In Phase I, the initial mass fraction of $^{18}$Ne (X($^{18}$Ne)) was extremely low at $7.5 \times 10^{-38}$, based on Lodders' data. This mass fraction decreased exponentially, approaching zero by the end of Phase III. Although, numerical residuals of order $10^{-39}$ remained detectable in Phase V.

The isotope $^{19}$Ne mass fraction (X($^{19}$Ne)) exhibited an initial increase during the first 2 Gyr of Phase I, after which it began to decay throughout the remainder of this phase. In Phase II, its mass fraction increased slightly, within a narrow range (1-8 $\times$ $10^{-35}$). Phase III was characterized by a continued decline of $^{19}$Ne, after which its mass fraction stabilized at trace levels during the final two phases.

The mass fraction of $^{20}$Ne (X($^{20}$Ne)) decreased gradually throughout Phase I and II, after which it remained nearly constant in Phases III and IV at $1.17 \times 10^{-3}$, showing no significant evolution during the red-giant onset and helium-burning phases. In Phase V, it underwent a rapid and substantial decline before stabilizing at approximately 10⁻¹⁴ by the end of the helium-exhaustion phase.

The mass fraction of $^{21}$Ne (X($^{21}$Ne)) increased during Phases I and II, continuing to rise through the first 100 Myr of Phase III. It then underwent a gradual decline, approaching near-zero values in the later phases.

The isotope $^{22}$Ne began with a mass fraction (X($^{22}$Ne)) of $7.3 \times 10^{-5}$, which decline until the end of Phase II. During Phases III and IV, its abundance remained nearly constant, while in Phase V it declines toward zero.

The evolution of all neon isotopes mass fraction with time during the five solar phases is shown in Appendix I. Table 2 present the mass fractions of all studied isotopes across the five evolutionary phases of the Sun.

Table 2: Mass fractions of the studied neon isotopes across the five evolutionary phases of the Sun.

| Element | Phase | Abundance | Comments |
|---|---|---|---|
| $^{18}$Ne | I | $7.5 \times 10^{-38}$ → $2.0 \times 10^{-39}$ | The mass fraction of $^{18}$Ne decreases rapidly during the main-sequence phase due to its very short half-life. Only numerical traces remain by the end of Phase I. |
| | II | $1.5 \times 10^{-38}$ → $1.2 \times 10^{-38}$ | The isotope remains at infinitesimal levels. Slight decrease is visible, but $^{18}$Ne is effectively absent. |
| | III | $\sim 10^{-35}$ → $\sim 10^{-287}$ | Rapid exponentials drop early in the phase. $^{18}$Ne is destroyed. |
| | IV | $\sim 10^{-270}$ → 0 | No production pathway exists under helium- |

| | | | burning conditions. |
|---|---|---|---|
| | V | ~$10^{-35}$ → 0 | Only computational noise remains |
| $^{19}$Ne | I | Peak ~$9.0 \times 10^{-34}$; end ~$1.8 \times 10^{-34}$ | $^{19}$Ne shows early rise then decay; short half-life keeps it at trace levels. |
| | II | $0.5 \times 10^{-35}$ → $9.3 \times 10^{-35}$ | Slight numerical increase but physically insignificant. |
| | III | ~$10^{-35}$ → ~$10^{-42}$ | Fully destroyed during rapid-growth/red-giant phase. |
| | IV | ~$10^{-40}$ → 0 | No sustainable production under helium-burning temperatures. |
| | V | ~$10^{-53}$ → 0 | All traces vanish during helium exhaustion. |
| $^{20}$Ne | I | $1.167425 \times 10^{-3}$ | $^{20}$Ne remains dominant; weak destruction causes small decline. |
| | II | $1.167429 \times 10^{-3}$ → $1.167421 \times 10^{-3}$ | Minor change; conditions too cool for significant processing. |
| | III | ≈$1.17 \times 10^{-3}$ | Abundance remains stable in early red-giant evolution. |
| | IV | $1.20 \times 10^{-3}$ → $1.18 \times 10^{-3}$ | Slight decrease due to weak destruction under helium burning. |
| | V | ~$10^{-8}$ → ~$10^{-14}$ | Rapid depletion from high-temperature photodisintegration. |
| $^{21}$Ne | I | $2.94 \times 10^{-6}$ → $3.06 \times 10^{-6}$ | $^{21}$Ne increases slowly during the main sequence. |
| | II | $2.940 \times 10^{-6}$ → $2.948 \times 10^{-6}$ | Small increase; stable isotope with mild response. |
| | III | ≈$1.05 \times 10^{-7}$ | Significant drop due to destruction channels. |
| | IV | ≈$1.05 \times 10^{-7}$ | Nearly constant during helium burning. |

| | | | | |
|---|---|---|---|---|
| | V | ~$10^{-23} \to 0$ | | Rapid decline during helium exhaustion. |
| $^{22}$Ne | I | $7.3\times10^{-5}$ $\to$ $1.0\times10^{-6}$ | | Strong exponential decline from the initial proto-solar abundance. |
| | II | $9.4\times10^{-5}$ $\to$ $4.7\times10^{-5}$ | | Smooth monotonic decline |
| | III | $\approx 5.5\times10^{-5}$ | | Nearly constant |
| | IV | $\approx 1.1\times10^{-5}$ | | Nearly constant and not changed. |
| | V | ~$10^{-6} \to 0$ | | Rapid decline during helium exhaustion. |

Abundances per nucleon were calculated for all five phases. Figure 7 illustrates the abundance per nucleon of $_{10}$Ne during Phase I, where it decreases to $5.6 \times 10^{-5}$ by the end of the phase and continues to decline into Phase II. In Phases III and IV, the abundance rises, reaching a maximum of $6.4 \times 10^{-5}$ at 80 Myr, before decreasing again by the end of both phases, as shown in Figure 8.

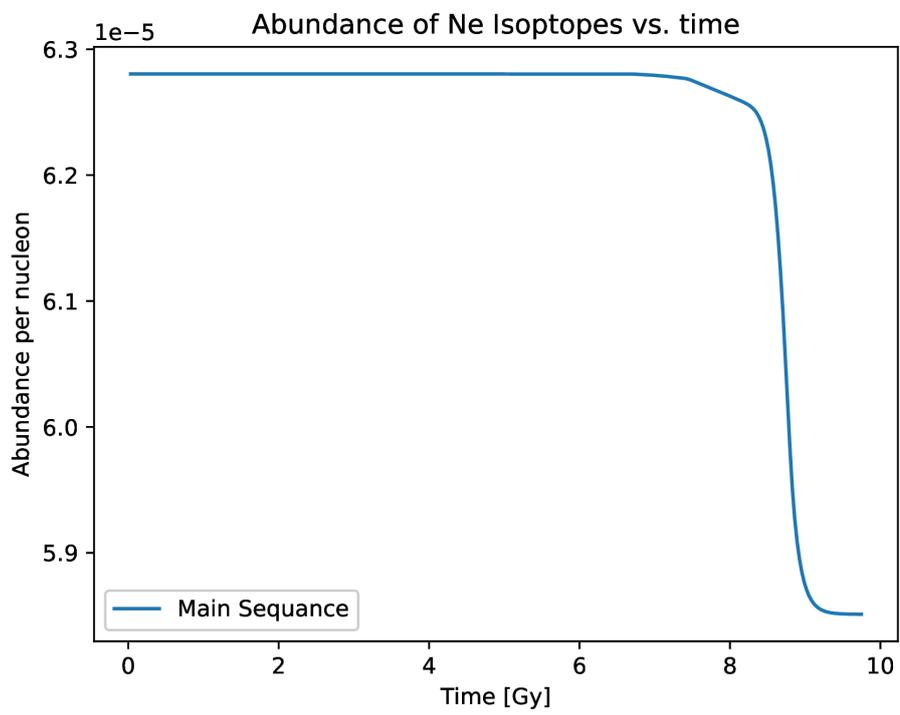

Fig. 7 Abundance per nucleon ($_{10}$Ne) evolution with time in Phase I.

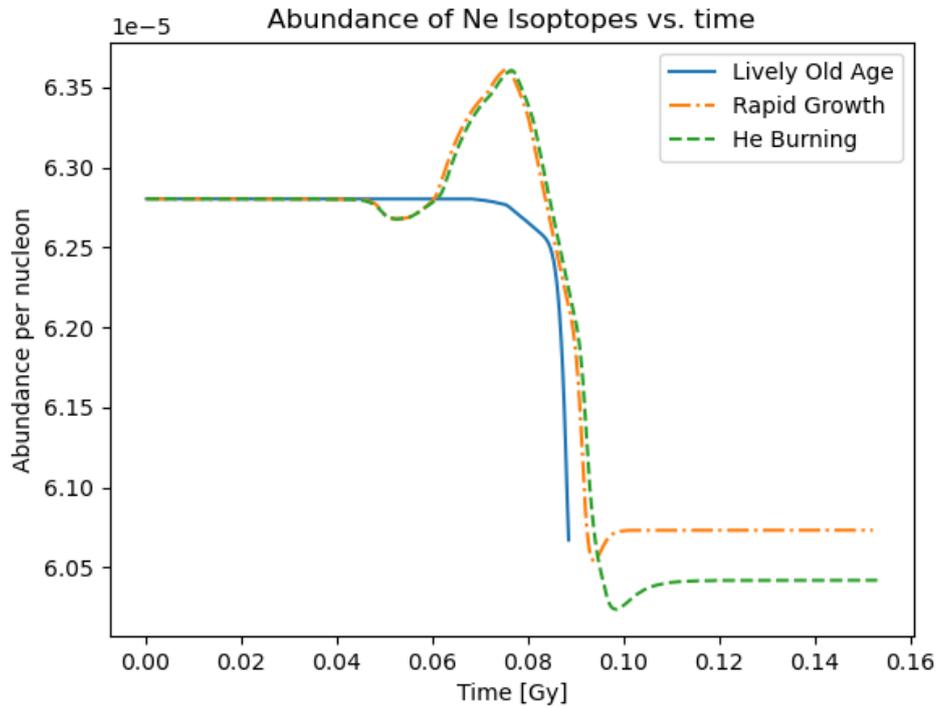

Fig. 8 Abundance per nucleon ($_{10}$Ne) evolution with time for II, III and IV phases.

The non-monotonic behavior seen in Figure 8 arises from the competition between production and destruction channels during these phases. In Phases III and IV, the prevailing temperature and density temporarily enhance the (n,γ) and (γ,n) pathways, leading to the a brief increase in abundance per nucleon. As the phase progresses, photodisintegration and neutron-capture channels become dominated, causing the abundance to decline again. This explains the "increase–decrease" trend shown in the figure.

The helium exhaustion phase (Phase V) initially exhibits a decline in the abundance per nucleon, reaching approximately $10^{-11}$. This is followed by a secondary increase before ultimately decreasing to nearly $10^{-15}$ by the end of the phase.

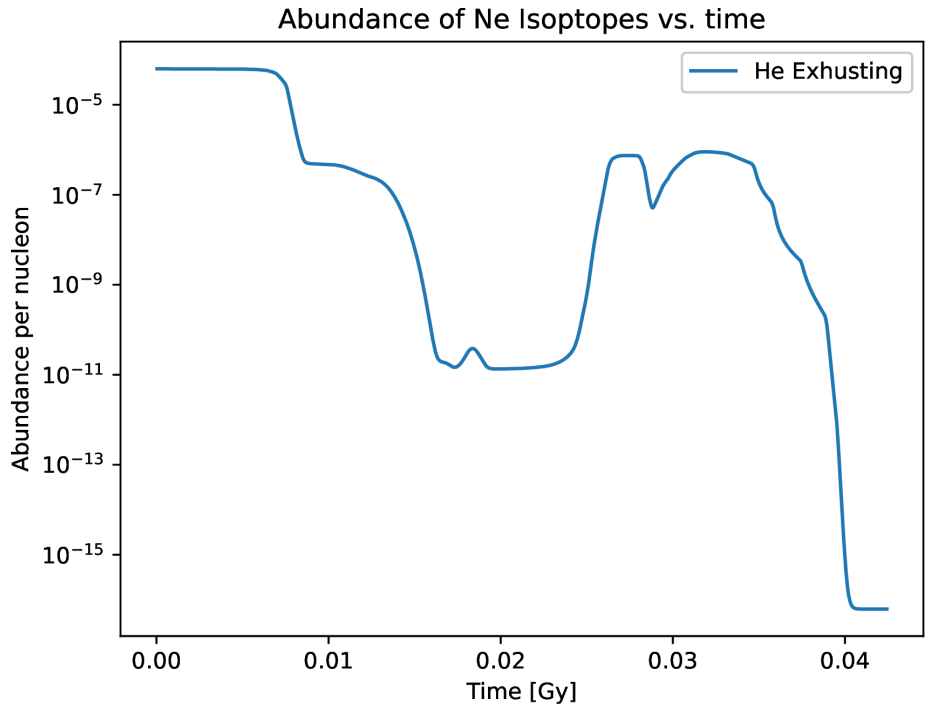

Fig. 9 Abundance per nucleon ($_{10}$Ne) evolution with time in Phase V.

At the beginning of Phase V (helium exhaustion), helium burning shuts down and the core begins to contract. This contraction raises the density and partially increases the temperature, initially enhancing destruction channels such as photodisintegration and neutron capture. As a result, the abundance per nucleon decreases sharply during the early part of the phase.

As the core continues to contract, the residual products of helium burning (mainly carbon, oxygen, and trace neutron sources) briefly activate weak production channels. During this period, reactions such as (n,γ) and limited (γ,n) cycling partially replenish some neon isotopes, leading to the secondary, temporary rise in abundance shown in Figure 9. This increase does not represent sustained production but rather a transient effect caused by the changing thermodynamic conditions near the termination of helium burning.

Toward the end of Phase V, the temperature and density approach the regime in which carbon-ignition conditions begin to develop, although the Sun itself will never ignite carbon. Nevertheless, the increasing core temperature strongly favors destruction pathways, particularly photodisintegration and α-capture on neon nuclei, while production channels become negligible due to the absence of helium and the weakness of remaining neutron sources. As a result, the abundance drops sharply again, approaching near-zero values by the end of the phase.

It is important to note that the mass fractions of all neon isotopes decrease over the five phases of the Sun's lifetime. The unstable isotopes $^{18}$Ne and $^{19}$Ne are effectively destroyed early in the evolution, whereas $^{20}$Ne and $^{22}$Ne experience modest production during the mid-phases (II-IV) before being depleted in the final phase. The stable isotope $^{20}$Ne maintains approximately the same order of magnitude through the end of Phase IV, after which it undergoes significant destruction during Phase V.

## 4. Conclusion

This study presents a phase-by-phase theoretical evaluation of the neon isotopes

$^{18}$Ne, $^{19}$Ne, $^{20}$Ne, $^{21}$Ne, and $^{22}$Ne throughout the five principal evolutionary stages of the Sun. Using the nucnet-tools nuclear reaction network under well-defined hydrostatic conditions, we quantified the evolution of each isotope in response to the temperature and density changes associated with hydrogen burning, old age, rapid growth and red giant, helium burning, and helium exhaustion.

The results reveal clear distinctions between the behaviors of unstable and stable isotopes. The short-lived isotopes $^{18}$Ne and $^{19}$Ne decay almost entirely during the hydrogen-burning phase and remain negligible thereafter. In contrast, the stable isotopes $^{20}$Ne and $^{22}$Ne dominate the neon composition throughout all phases. Specifically, $^{20}$Ne maintains the highest mass fraction over the Sun's lifetime, exhibiting only modest depletion until a sharp decline during helium exhaustion. The intermediate isotope $^{21}$Ne increases slightly during the hydrogen burning phase, remains nearly constant during helium burning, and then undergoes rapid destruction at the onset of core contraction.

Although several of the predicted mass fractions become extremely small at late evolutionary stages, these near-zero values remain physically meaningful. They reflect the expected behavior of the nuclear reaction network, in which rapid decay, efficient destruction channels, and the absence of sustained production pathways drive certain isotopes to negligible levels. Reporting these small abundances is essential for distinguishing which isotopes survive over stellar timescales and which are eliminated, thereby capturing the full nucleosynthetic evolution of neon in solar-type stars.

By providing quantitative abundance predictions for the helium-burning and helium-exhaustion phases, stages for which published neon-isotope data remain limited, this work offers a unified reference model for neon nucleosynthesis in solar-type stars. These results may support future studies involving stellar evolution modeling, isotopic diagnostics of the solar interior, and comparisons with observational or meteoritic constraints.

**Acknowledgments**

I would like to express my thanks and gratitude to the Sharjah Academy for Astronomy, Space Sciences and Technology (SAASST) for facilitating this project.

**Appendix A. A phase-by-phase evolution of neon isotopes mass fractions during the full lifetime of the Sun.**

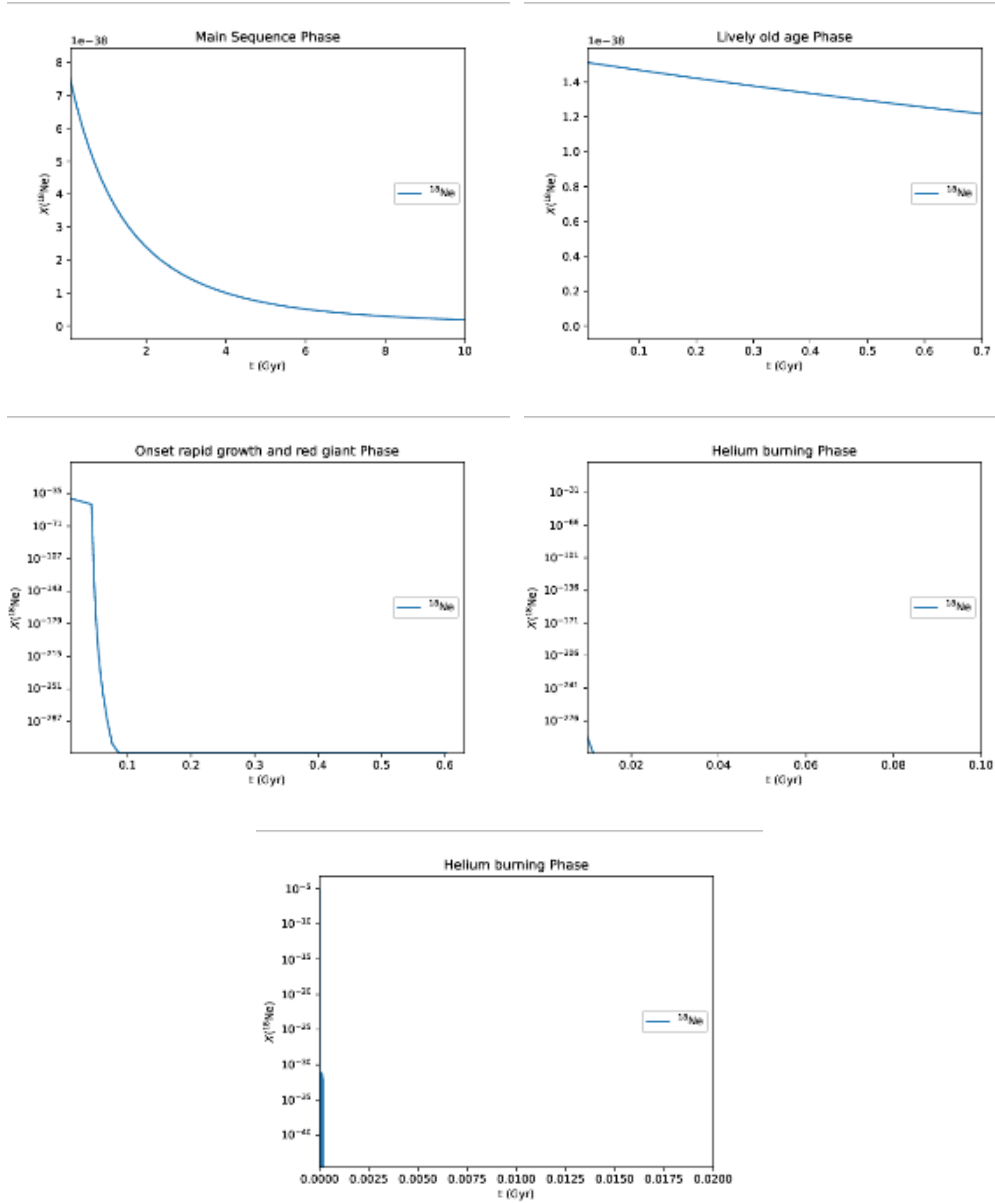

Figure A1: The evolution of the $^{18}$Ne mass fraction ($X(^{18}Ne)$) across the five solar evolutionary phases. Panels are ordered from top to bottom and left to right from Phase I to V. Each panel displays the abundance of evolution corresponding to its respective phase.

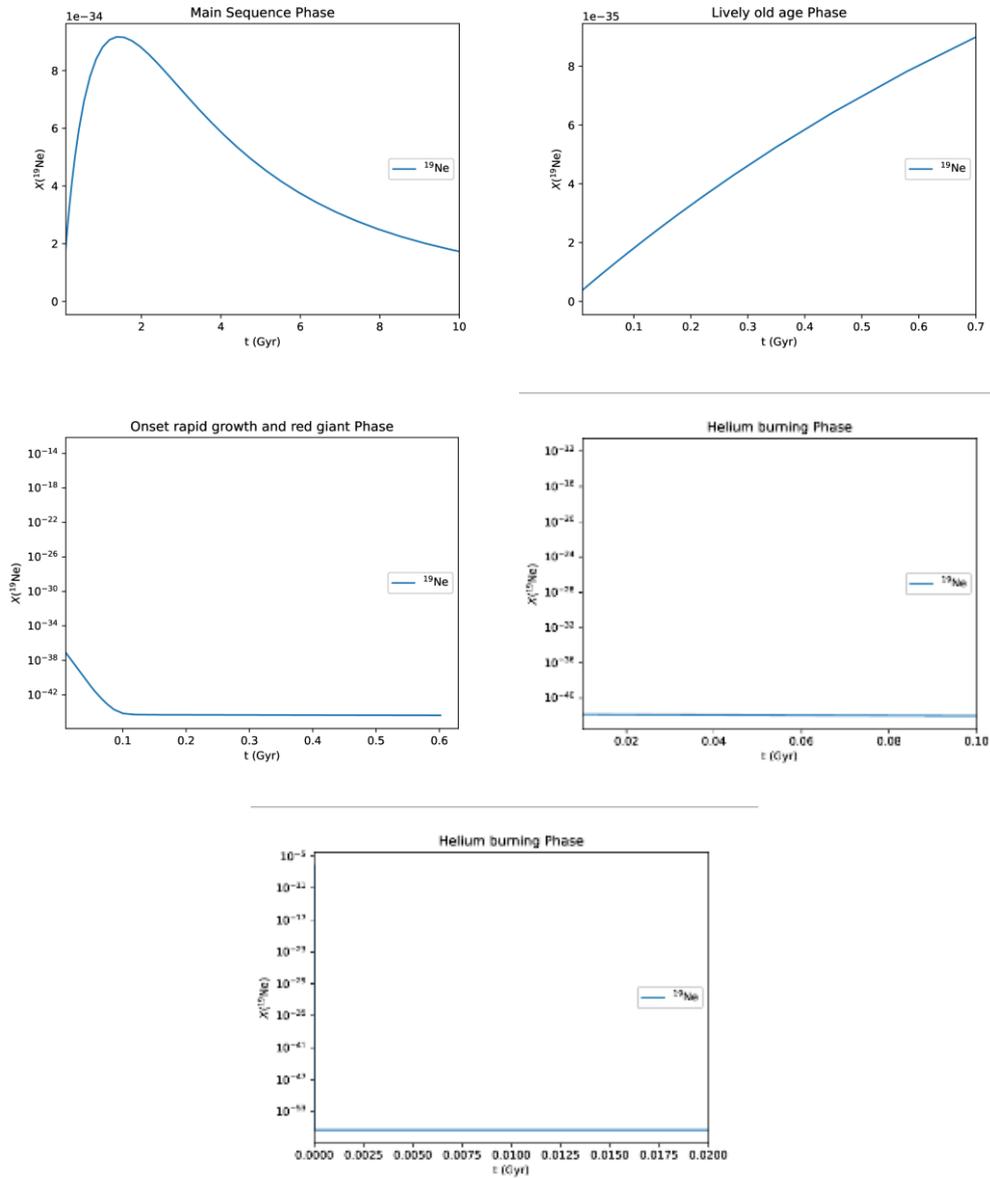

Figure A2: The evolution of the $^{19}$Ne mass fraction ($X(^{19}Ne)$) across the five solar evolutionary phases. Panels are ordered from top to bottom and left to right from Phase I to V. Each panel displays the abundance of evolution corresponding to its respective phase.

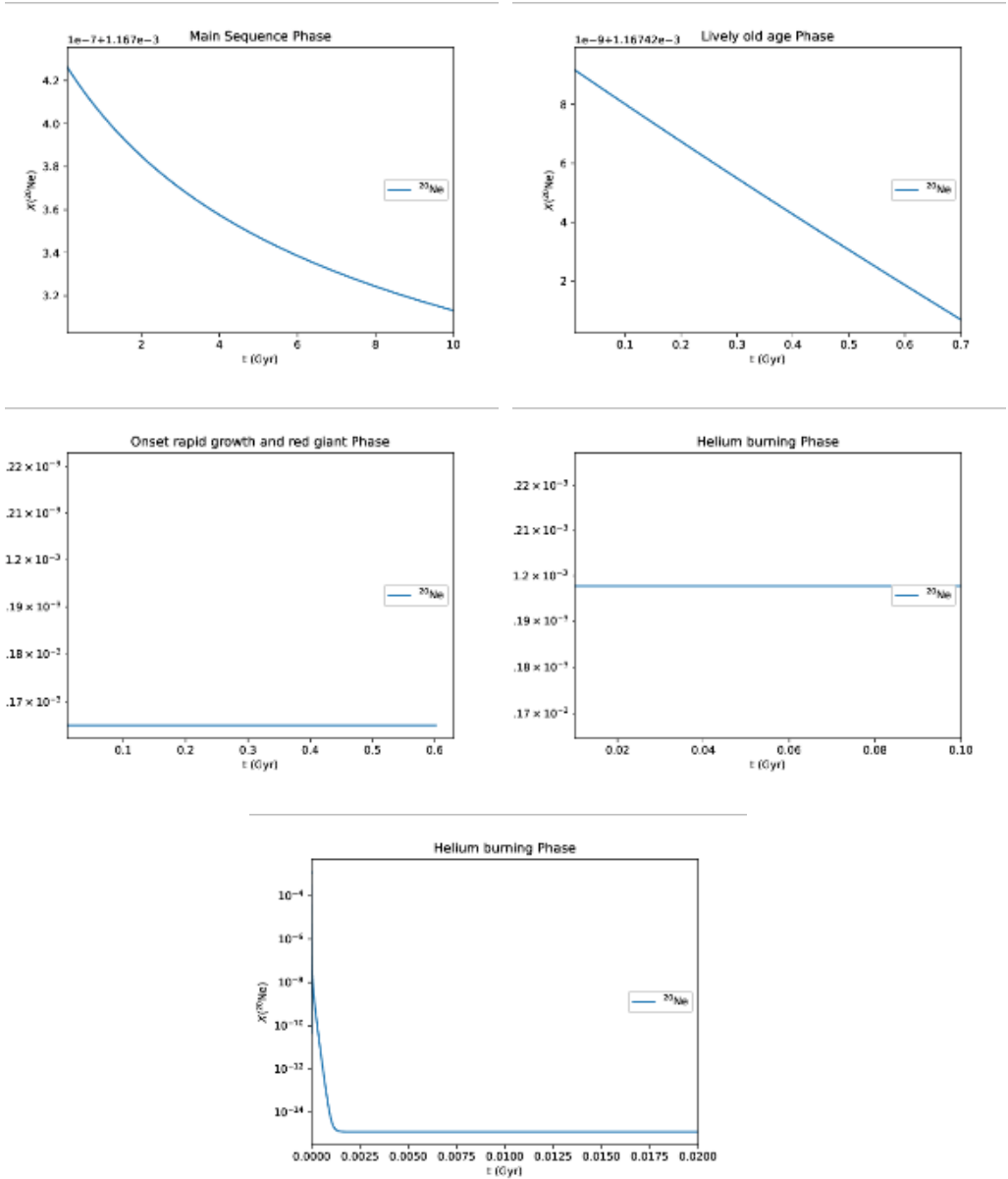

Figure A3: The evolution of the $^{20}$Ne mass fraction ($X(^{20}Ne)$) across the five solar evolutionary phases. Panels are ordered from top to bottom and left to right from Phase I to V. Each panel displays the abundance of evolution corresponding to its respective phase.

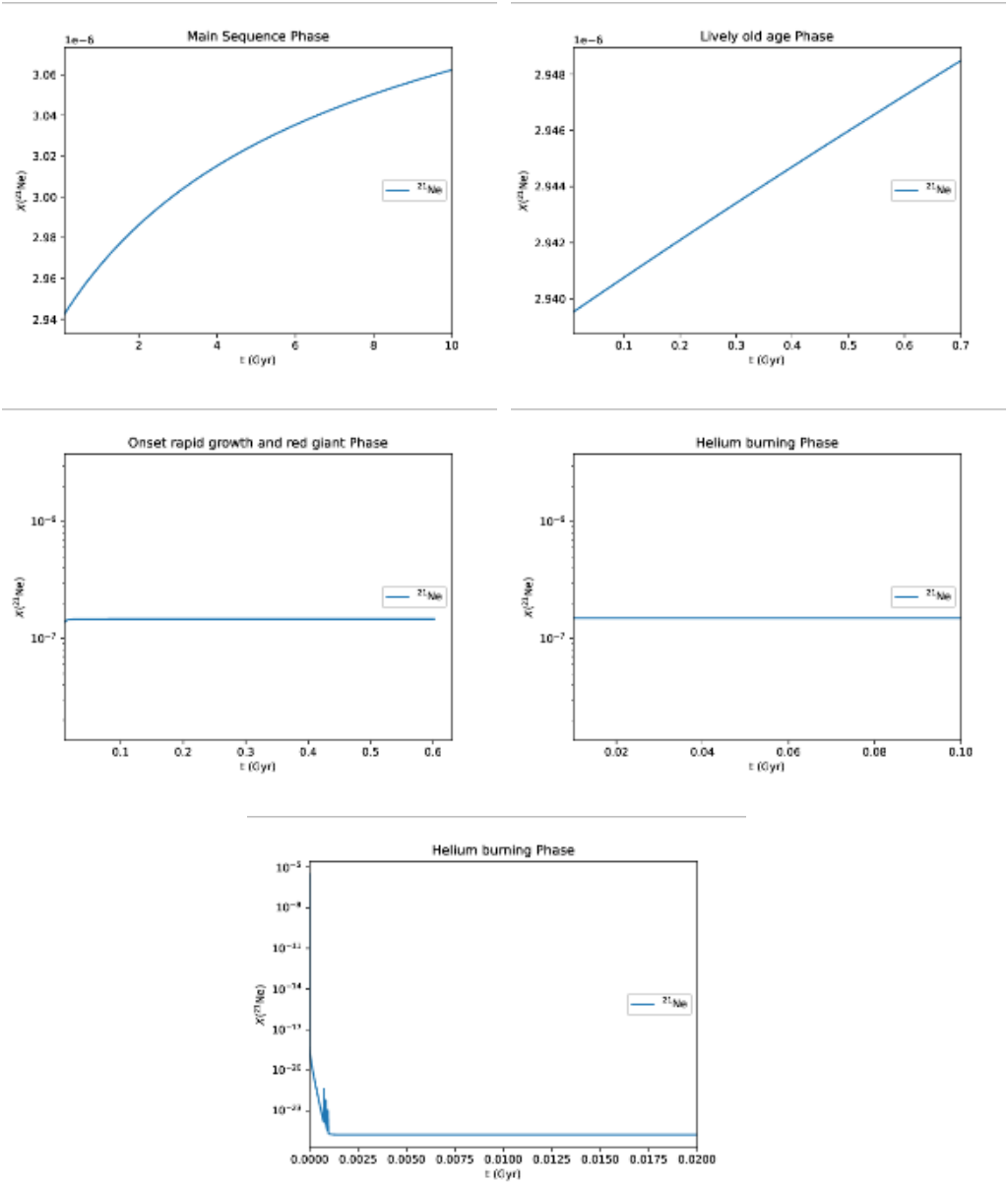

Figure A4: The evolution of the $^{21}$Ne mass fraction ($X(^{21}Ne)$) across the five solar evolutionary phases. Panels are ordered from top to bottom and left to right from Phase I to V. Each panel displays the abundance of evolution corresponding to its respective phase.

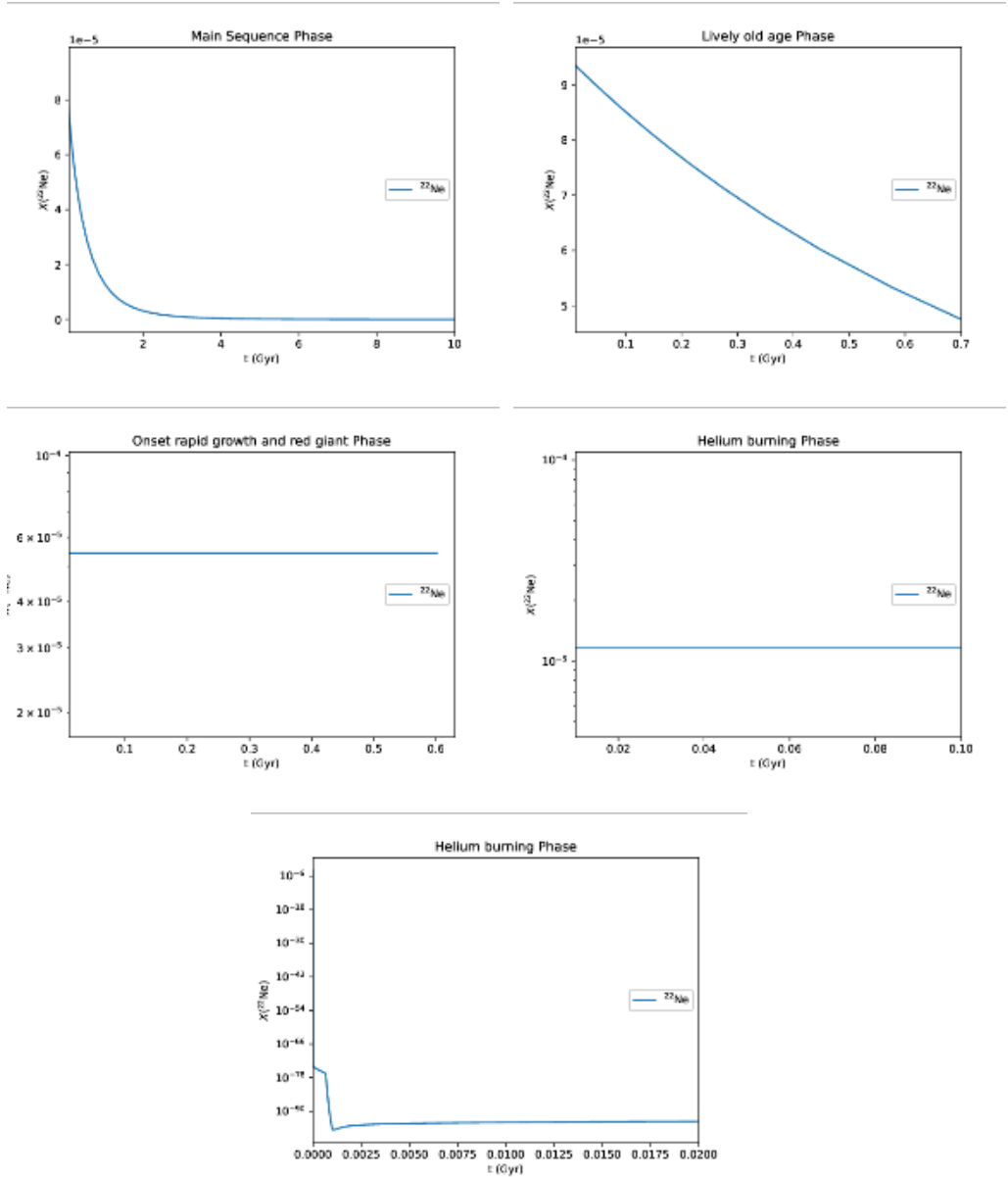

**Figure A5:** The evolution of the $^{22}$Ne mass fraction ($X(^{22}Ne)$) across the five solar evolutionary phases. Panels are ordered from top to bottom and left to right from Phase I to V. Each panel displays the abundance of evolution corresponding to its respective phase.